\documentclass[fleqn,10pt]{wlscirep}
\usepackage[utf8]{inputenc}
\usepackage[T1]{fontenc}
\usepackage{hyperref}

\title{Collective dynamics of stock market efficiency}
\author[1]{Luiz G. A. Alves}
\author[2]{Higor Y. D. Sigaki}
\author[3,4,5,*]{Matja{\v z} Perc}
\author[2]{Haroldo V. Ribeiro}

\affil[1]{Department of Chemical and Biological Engineering, Northwestern University, Evanston, IL 60208, U.S.A.}
\affil[2]{Departamento de F\'isica, Universidade Estadual de Maring\'a, Maring\'a, PR 87020-900, Brazil}
\affil[3]{Faculty of Natural Sciences and Mathematics, University of Maribor, Koro{\v s}ka cesta 160, 2000 Maribor, Slovenia}
\affil[4]{Department of Medical Research, China Medical University Hospital, China Medical University, Taichung, Taiwan}
\affil[5]{Complexity Science Hub Vienna, Josefst{\"a}dterstra{\ss}e 39, 1080 Vienna, Austria}

\affil[*]{matjaz.perc@gmail.com}

\begin{abstract}
Summarized by the efficient market hypothesis, the idea that stock prices fully reflect all available information is always confronted with the behavior of real-world markets. While there is plenty of evidence indicating and quantifying the efficiency of stock markets, most studies assume this efficiency to be constant over time so that its dynamical and collective aspects remain poorly understood. Here we define the time-varying efficiency of stock markets by calculating the permutation entropy within sliding time-windows of log-returns of stock market indices. We show that major world stock markets can be hierarchically classified into several groups that display similar long-term efficiency profiles. However, we also show that efficiency ranks and clusters of markets with similar trends are only stable for a few months at a time. We thus propose a network representation of stock markets that aggregates their short-term efficiency patterns into a global and coherent picture. We find this financial network to be strongly entangled while also having a modular structure that consists of two distinct groups of stock markets. Our results suggest that stock market efficiency is a collective phenomenon that can drive its operation at a high level of informational efficiency, but also places the entire system under risk of failure.
\end{abstract}
\begin{document}

\flushbottom
\maketitle

\thispagestyle{empty}

\section*{Introduction}

The efficient market hypothesis is a paradigm in financial economics and a widespread belief among stock market agents~\cite{cootner1964random,fama1970efficient}. The hypothesis states that asset prices fully reflect all available information in an ideally efficient stock market~\cite{fama1965behavior}. As new information about individual stocks or the stock market becomes available, it is immediately priced by the stock market agents producing new values for the assets~\cite{malkiel2003efficient}. This idea is frequently associated with the concept of a random walk, in which future price changes represent random departures from previous prices. Accordingly, attempts to predict future prices in informationally efficient markets are likely no better than a random guess. The efficient market hypothesis thus imposes substantial limitations to financial trading and makes it impossible to surpass efficient markets via arbitrage or trading strategy~\cite{malkiel2003efficient}.

While the efficient market hypothesis is a crucial concept in financial economics, real stock markets are not ideally efficient~\cite{sornette2017stock}. Stock prices in real-world markets can become auto-correlated during short-term periods~\cite{stanley2001similarities}, corroborating the more holistic idea that making short-term predictions and arbitrage are possible. Another evidence that stock markets are not ideally efficient are the non-Gaussian fluctuations (fat-tailed distributions) of the log-returns of asset prices~\cite{gopikrishnan1999scaling,mantegna1995scaling}, the difficulty of simple random walk models in predicting stock market crashes~\cite{johansen2000crashes}, and the existence of successful trading strategies~\cite{preis2013quantifying}. Furthermore, while it is believed that efficient markets prevent economic bubbles and crashes~\cite{preis2011bubble}, it is precisely the long-range correlations present in such events that make them more predictable~\cite{sornette2002predictability}.

Whether the lack of efficiency in stock markets is an opportunity for profit, or whether it represents a systemic risk for financial systems that must be fixed -- fact is that the efficient market hypothesis is still an ubiquitous concept among economic agents and academics working with financial data and models. This interest translates into a wealth of works that try to quantify the degree of efficiency in different stock markets~\cite{zunino2007inefficiency,zunino2008multifractal,zunino2009forbidden,zunino2010complexity,zunino2012efficiency,szarek2020long,wang2020long,rocha2020evidence,sanchez2020testing} and new forms of investment such as cryptocurrencies~\cite{urquhart2016inefficiency,bariviera2017some,zhang2018inefficiency,bariviera2017inefficiency, nadarajah2017inefficiency,tiwari2018informational,bariviera2018analysis,alvarez2018long,sigaki2019clustering,dimitrova2019some}. Despite the increasing engagement in better understating different facets of the efficient market hypothesis, a large part of these studies assume that market efficiency remains unchanged throughout the investigated period. Thus, the more realistic possibility of having stock markets with time‐varying degree of efficiency remains considerably less explored. This is indeed regrettable since time-dependent efficiency measures allow us to probe for collective motions in the evolution of market efficiency, as well as to quantify the stability of different efficiency ranks in stock markets.

Here we investigate the dynamical behavior of the efficiency of 43 major world stock markets during the past 20 years. We use a physics-inspired approach for defining a time‐varying efficiency from log-returns of stock market indices. Specifically, we define the time-dependent efficiency of a stock market as the permutation entropy~\cite{bandt2002permutation} calculated within sliding time-windows of log-returns. Because the permutation entropy estimates the degree of randomness in a time series, we expect this measure to be close to one when a market is in an informationally efficient state. Conversely, values smaller than one indicate a departure from the random behavior such that the smaller the entropy value, the less efficient the stock market.

Our research shows that major world stock markets can be hierarchically classified into several clusters according to the similarity in the long-term evolution of their efficiency degrees. However, we find that the efficiency ranks of stock markets and clusters of markets with similar efficiency trends are only stable during short periods of time, typically lasting no longer than a few months. This result indicates that grouping markets by their long-term behavior may hide important aspects of their interactions with other markets. To unveil such features, we use a dynamical clustering approach that is able to find groups of stock markets with similar efficiency patterns within a time window. By using these time-varying clusters, we construct a weighted network where nodes represent stock markets, edges indicate markets that appear together in the same group at least once, and edge weights are proportional to the number of times a pair of markets appears in the same cluster. This network allows us to identify the most influential markets, as well as their modular structure, which comprises two distinct market groups that have similar efficiency trends. We further observe that the obtained network of stock markets is very dense and entangled. Thus, we conclude that the efficiency dynamics of stock markets is a collective phenomenon that can drive the entire financial system to operate at a very high level of informational efficiency -- a level that might be unattainable by any other means -- but that also places the entire system under continuous and systemic risk of failure. 

\section*{Results}

\begin{figure}[!ht]
\centering
\includegraphics[width=0.8\linewidth]{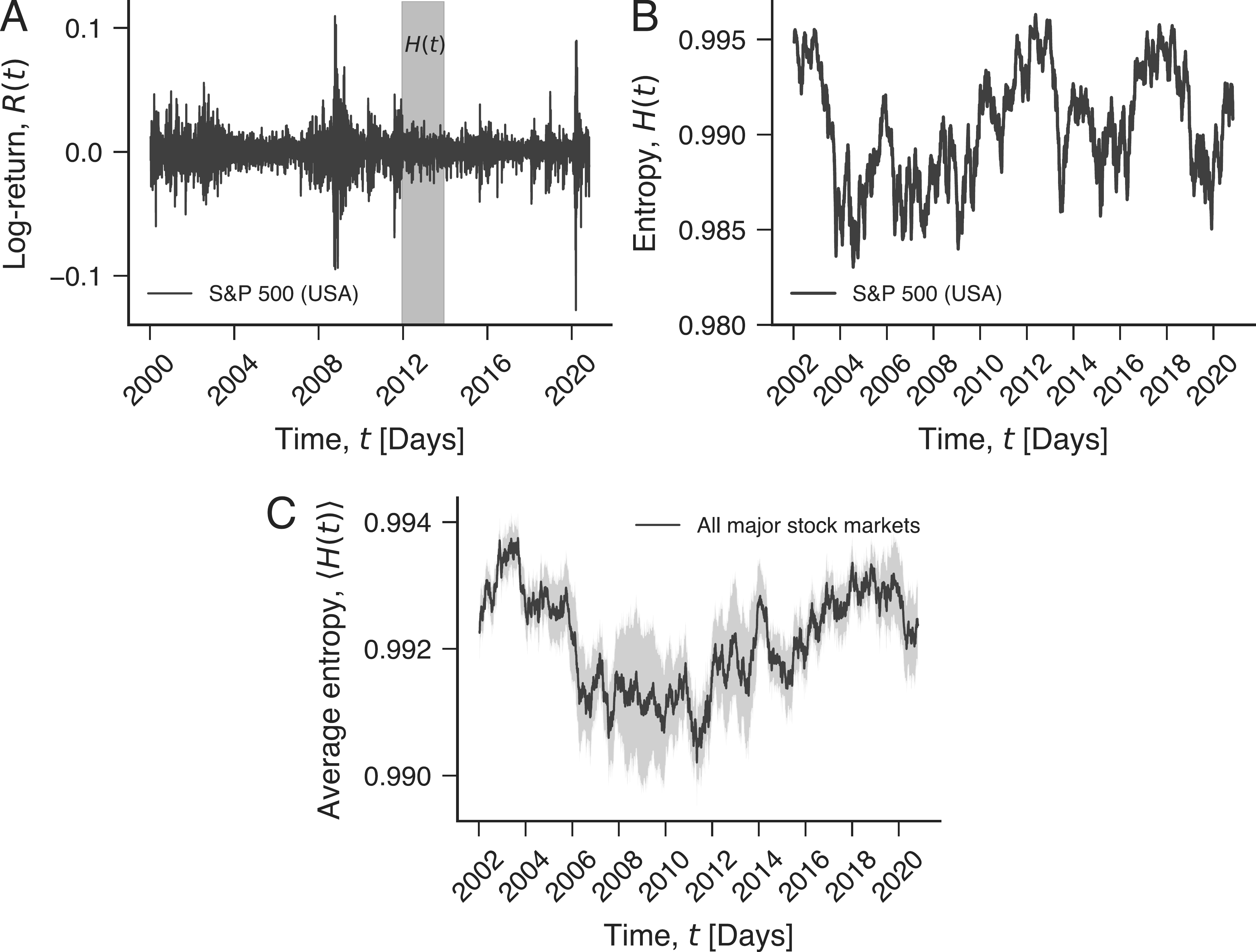}
\caption{Defining the informational efficiency of stock markets with permutation entropy. (A) Log-return time series $R(t)$ of the S\&P 500 daily closing prices from January 1, 2000 to October 31, 2020. The shaded area illustrates a 500-day sliding window (two years of stock market operation) used for calculating the permutation entropy $H(t)$. (B) Time evolution of the permutation entropy $H(t)$ with embedding dimension $d=4$ (Methods Section for details) of the S\&P 500 index. (C) Time evolution of the average value of the efficiency $\langle H(t)\rangle$ of all 43 stock markets in our study (shaded band represents the standard error of the mean).}
\label{fig:1}
\end{figure}

Our results are based on the daily closing indices of 43 major world stock markets from January 2000 to October 2020 (Methods Section for details and Table~\ref{stab:1} for the list of market indices). These historical time series represent the stock market indices after adjustments for all applicable splits and dividend distributions (daily adjusted closing prices). From these time series, we estimate the log-returns $R(t)$ (Methods Section for details) of each market index where $t$ stands for the closing date. Figure~\ref{fig:1}A illustrates the time evolution of $R(t)$ for the S\&P 500 index (an influential indicator of the USA equity market), while Fig.~\ref{sfig:1} shows the behavior of $R(t)$ for all stock markets in our study. Next, we sample the log-return series with a 500-day sliding window (shaded gray in Fig.~\ref{fig:1}), roughly corresponding to two years of economic activity. The sliding window moves with daily steps, and for every step, we estimate the normalized permutation entropy $H(t)$~\cite{bandt2002permutation}. This procedure creates a time series of the permutation entropy $H(t)$ for each stock market, as shown in Fig.~\ref{fig:1}B for the S\&P 500 (see Fig.~\ref{sfig:2} for all markets). 

As detailed in the Methods Section, the permutation entropy is estimated from ordering patterns among consecutive values of $R(t)$ and quantifies the degree of randomness in the occurrence of these patterns. We expect an entirely regular series to have $H\approx0$, while a completely random series displays $H\approx1$. Thus, the higher the value of $H(t)$, the more random the log-return series is around time $t$, and so the more informationally efficient is the stock market at that particular time. Conversely, a decrease in $H(t)$ indicates the emergence of a more regular (and possibly more predictable) behavior of the log-return series, and thus a less efficient period of the stock market. We have also estimated the average behavior of the efficiency  $\langle H(t) \rangle$ over all stock markets. Figure~\ref{fig:1}C shows that the aggregate behavior is smoother than the behavior observed for individual stock markets and appears to reflect major financial events such as the ``global financial crisis'' (2007-2008), as around this period $\langle H(t) \rangle$ displays lower values of entropy. 

\begin{figure}[!ht]
\centering
\includegraphics[width=0.99\linewidth]{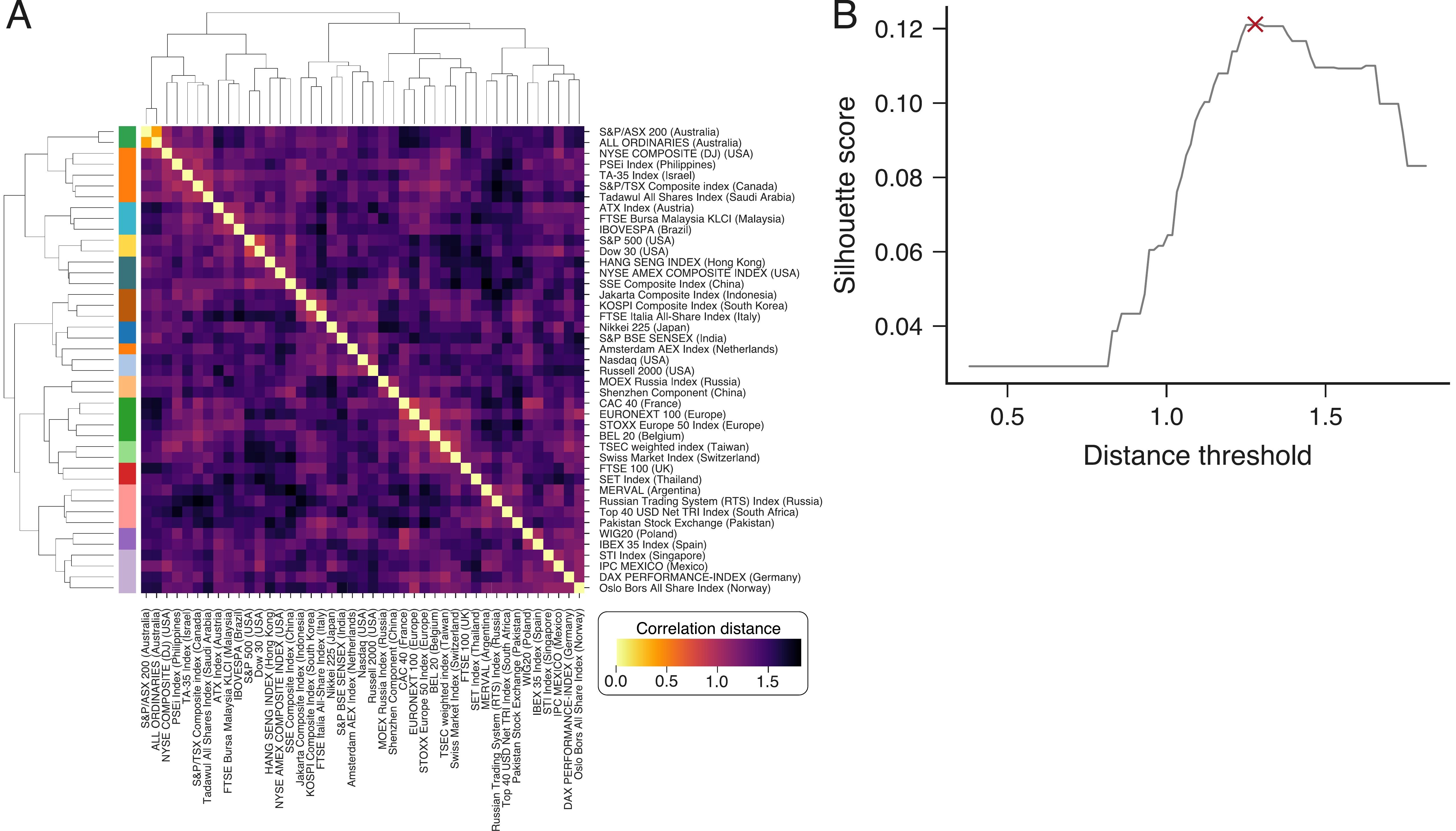}
\caption{Long-term hierarchical organization of efficiency patterns of major world stock markets. (A) Matrix plot of the correlation distance among all pairs of entropy time series of stock markets. The dendrograms that go along this matrix show the hierarchical clustering result based on Ward's minimum variance method. (B) Silhouette score calculated from clusters obtained by cutting the dendrogram at different threshold distances. The colored squares located below the dendrogram branches in panel A indicate the 16 clusters obtained by cutting the dendrogram at the threshold distance that maximizes the silhouette score (indicated by the red cross).}
\label{fig:2}
\end{figure}

Investors' behavior tends to synchronize during stock market crashes~\cite{sornette2017stock}, and investment strategies can propagate shocks through financial networks and lead to the emergence of strong correlations among financial markets~\cite{delpini2019systemic}. Similarly, we expect these collective behaviors to strongly affect the efficiency dynamics of stock markets and produce joint movements in $H(t)$ that may organize markets into hierarchical structures with similar efficiency trends. To investigate this possibility, we first estimate the correlation distance of the efficiency time series among all pairs of markets (Methods Section for details), creating the correlation distance matrix shown in Fig.~\ref{fig:2}A. Next, we use Ward's minimum variance method (Methods Section for details) to build up a dendrogram representation of the distance matrix, which is also depicted in Fig.~\ref{fig:2}A. Our results indicate that stock markets form a hierarchical structure regarding their long-term efficiency evolution. However, we do not observe large clustering structures in the distance matrix. Indeed, by determining the number of clusters by maximizing the silhouette score (as shown in Fig.~\ref{fig:2}B, see Methods Section for details), we end up with 16 clusters in which 15 consist of only a few markets (the largest cluster consists of 5 markets) and 1 with only a single market. These groups of markets exhibit similar long-term temporal profiles of $H(t)$ (see Fig.~\ref{sfig:3}), but this global analysis does not capture short-term movements of $H(t)$ among stock markets. 

\begin{figure}[!ht]
\centering
\includegraphics[width=0.99\linewidth]{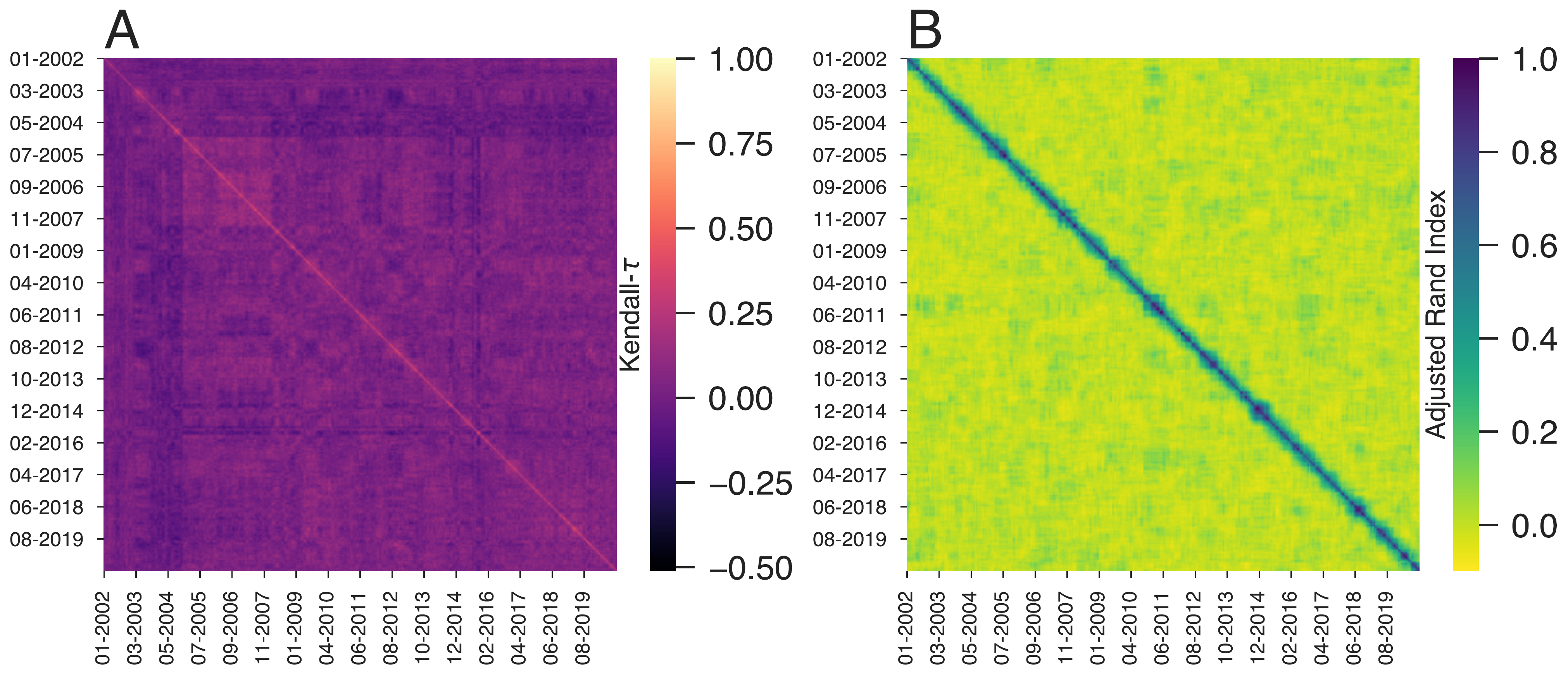}
\caption{Short-term stability of efficiency rankings and clusterings of stock markets. (A) Matrix plot of the Kendall rank correlation coefficient (Kendal-$\tau$) among all pairs of efficiency ranks of stock markets calculated within a one-year sliding window. We note the formation of small diagonal blocks with widths of one or two months, indicating that the efficiency ranks change over time. (B) Matrix plot of the adjusted rand index estimating the agreement among clustering of markets with similar temporal efficiency profiles at different time-windows. We also observe the existence of small diagonal blocks with widths of approximately four months, showing that groups with similar efficiency profiles do not remain stable during long periods.}
\label{fig:3}
\end{figure}

To start investigating short-term collective behavior in the efficiency of stock markets, we sample the time series of $H(t)$ with a one-year sliding window and create a ranking of markets based on the average efficiency within each time-window. Next, we investigate the stability of these efficiency rankings by estimating the Kendall rank correlation coefficient (Kendal-$\tau$) among all possible pairs of time-windows. This analysis results in the correlation matrix shown in Fig.~\ref{fig:3}A, where the rows and columns represent the last date of each time-window. By definition, the diagonal elements of this matrix are unitary (that is, the efficiency rank in a given time window is perfectly correlated to itself). Values along a given row or column indicate how similar is the efficiency rank of that particular date with past and future rankings. Thus, we would expect large diagonal blocks with high Kendal-$\tau$ values if these efficiency ranks were stable over long periods. However, we observe small diagonal blocks with about one-month width, indicating that these efficiency ranks are stable only under very short periods.

We have also calculated the correlation distance of the efficiency $H(t)$ among all pairs of markets for each time-window and applied the same clustering approach used for the entire time series of $H(t)$ (that is, Ward's minimum variance dendrogram segmented at the maximum silhouette score). This approach produces groups of markets with similar efficiency evolution in each time-window. We compare the temporal stability of these groups by estimating the adjusted rand index among all clustering results in each time-window. This coefficient measures the agreement between two clusterings by counting pairs of elements assigned in the same group while controlling for the overlap expected by chance. Values of adjusted rand index around zero indicate that the two clusterings are no more similar than random partitions, while 1 indicates perfect agreement between them. Figure~\ref{fig:3}B shows the matrix plot of the adjusted rand index for all pairs of clustering results. The small block-diagonal structures of the adjusted rand index matrix indicate that groups of markets with similar $H(t)$ profiles remain stable only for approximately four months.

The results of Fig.~\ref{fig:3} demonstrate that short-term collective patterns in the efficiency evolution of stock markets change with time and are different from those obtained at long time scales. The dynamical patterns we have found indicate that simple partitions are not enough to capture the complex interactions among financial markets and motivate the use of a different approach that considers these entangled interactions. To do so, we propose to build a complex network where nodes represent the stock markets, and links indicate two markets that clustered together at least once over time. We also assume that the connection between two stock markets are weighted by the number of times those particular markets are grouped in the same cluster. This representation allows us to aggregate the short-term information into a global and coherent picture where stock markets whose efficiency dynamics are correlated during some period are connected; furthermore, the strengths of these connections indicate the intensity of the interactions among the markets. 

\begin{figure}[!t]
\centering
\includegraphics[width=0.756\linewidth]{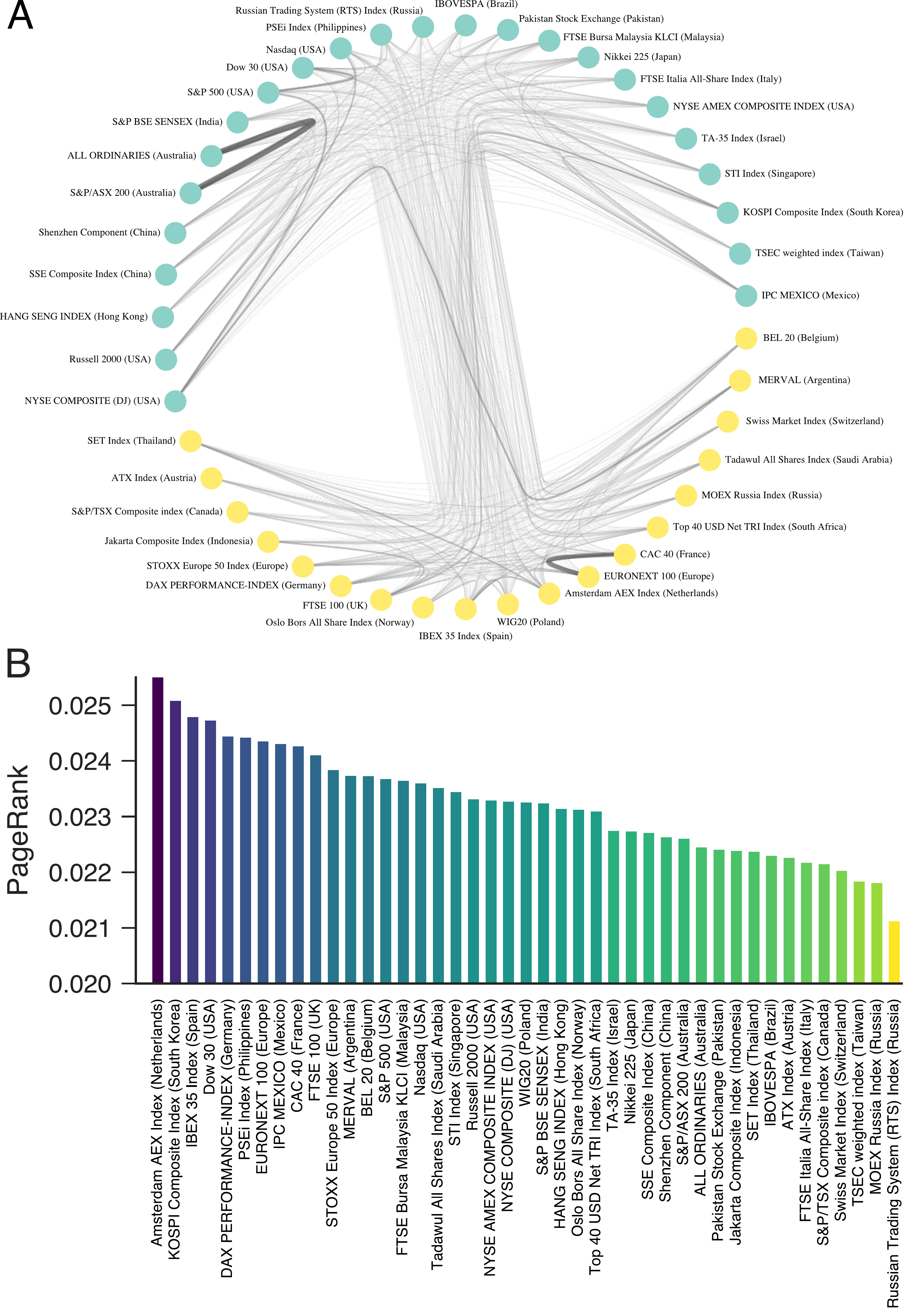}
\caption{Financial network of stock markets exhibiting similar short-term trends of efficiency. (A) Nodes represent stock markets, and links are drawn among markets appearing at least once in the same cluster regarding the short-term similarity in their evolution of efficiency. These links are further weighted by the number of times pairs of markets are grouped in the same cluster. By using the nested stochastic block model (Methods Section for details), we identify two network modules indicated by the different node colors. In this visualization edge widths are proportional to the weights. (B) Centrality ranking based on PageRank indicating the most influential stock markets for the efficiency dynamics of the stock market indices.}
\label{fig:4}
\end{figure}

Figure~\ref{fig:4}A shows this network representation for the 43 stock markets in our study. We observe that this complex network forms a complete graph as it presents all possible connections among all stock markets. Thus, world stock markets are strongly globalized regarding their efficiency such that simultaneous periods of high or low efficiency may involve a large number of markets. This result suggests the existence of systemic risk for the ``spreading'' of low-efficient states but, at the same time, it indicates that high-efficient states can also globally emerge. Although the density of this financial network is maximum, the interactions among the stock markets are not uniformly distributed. The Gini coefficient of the edge weights is $0.18$ (on a scale where zero means perfect equality and one maximal inequality) and strengthens the idea that some markets may have a higher impact on the efficiency dynamics of the entire system. Figure~\ref{fig:4}B shows a centrality ranking based on PageRank~\cite{brin2012reprint} where the Amsterdam AEX Index (Netherlands) and KOSPI Composite Index (South Korea) emerge as the most influential markets, while the two Russian index (MOEX Russia Index, Russian Trading System (RTS) Index) are the less influential for the efficiency dynamics of the world stock markets.

The inequality in the edge weights distribution also suggests that the financial network of Fig.~\ref{fig:4}A may have a modular structure in which groups of markets are more similar among themselves than with other groups. To probe for this possible modular structure, we use the stochastic block model approach~\cite{holland1983stochastic,funke2019stochastic,peixoto2014hierarchical}. As detailed in Methods Section, we have fitted different stochastic block models to our network data, finding that the nested stochastic block model without degree correction is the best model description for our network. This model yields the two network modules represented by the different node colors in Fig~\ref{fig:4}A. 

The largest network module (with 24 indices) comprises markets from USA (6 indices), Asia-Pacific countries (14 indices), and 4 other markets from Brazil, Italy, Israel, and Mexico. The second major module includes the remaining 19 indices: 12 from Europe, and other markets from Argentina, Canada, Indonesia, Russia, Saudi Arabia, South Africa, and Thailand.  Despite the existence of many exceptions, the geographical distance appears to play a role in these partitions. However, more important than understanding the particularities of each module is the emergence of this modular structure. Although the associations among stock markets are quite entangled, this modular structure suggests that some groups of markets are more similar to each other such that low- or high-efficiency states are more likely to encompass these modules.

\section*{Discussion}

We have presented an investigation of dynamical efficiency patterns of 43 major world stock markets during the past 20 years. To do so, we have relied on a physics-inspired approach in which the efficiency of a stock market at a particular time is defined by estimating the permutation entropy within sliding time-windows of log-returns of stock market indices. Our results indicate that stock markets can be hierarchically organized into groups of markets having similar long-term efficiency trends. However, we have also found that these long-term clusters are not enough to fully understand the collective dynamics of market efficiency. Indeed, our research has revealed that short-term collective patterns in the evolution of efficiency vary with time, and are different from those at longer time scales. We have observed that efficiency ranks of stock markets are stable only during relatively short periods of time, no longer than a month or two. Similarly, the clustering of markets with similar efficiency profiles is stable only for approximately four months. 

Because of these facts, we have proposed a complex network representation where nodes are stock markets, and connections among them indicate markets that clustered together at least once during the 20 years of our data. We have further considered that links between a pair of markets are weighted by the number of times they appear in the same cluster regarding their short-term efficiency dynamics. Our results show that this financial network is fully-connected, indicating that the efficiency of stock markets are strongly entangled and globalized. Previous works have already demonstrated that systemic failures in stock markets emerge from a synchronization process~\cite{johansen2000crashes} that takes place in social networks~\cite{castellano2009statistical,perc2017statistical} and may cause bubbles that eventually burst~\cite{sornette2002predictability}. Studies also suggest the existence of strong correlations between fund investment strategies as a cause of systemic risks and shock propagation~\cite{delpini2019systemic}. In this context, the intricate financial network uncovered by our work suggests a systemic risk for the ``spreading'' of low-efficiency states but, at the same time, indicates that high-efficiency states can also emerge at a global level. Although our financial network forms a complete graph, the link weights among stock markets are not uniformly distributed, allowing us to identify the most influential markets. Furthermore, we have found this financial network to have a modular structure that comprises two market groups whose efficiency dynamics is more similar within the groups as it is with markets outside of the groups. Therefore, despite stock markets being quite entangled in terms of their efficiency profiles, the modular structure indicates that low-efficient and high-efficient states are more likely to emerge within these groups.

Because efficiency in stock markets can be an opportunity for profit as well as an early signal of an impending financial crisis, it would be interesting to incorporate our approach into models that try to predict stock market prices to measure financial transaction risks. For instance, researches used \textit{Google Trends} to inform investors when to buy or sell stocks and found that their strategy was 326\% better than the traditional buy-and-hold strategy~\cite{preis2013quantifying}. Despite the impressive improvement of this strategy, it does not inform the buying or selling volumes, or the risks that are involved in the proposed transactions. Thus, quantifying the time dependence of the stock efficiency may help economic agents to quantify their transaction risks.

Naturally, our study has its limitations. The correlations among stock markets expressed in our network representation do not carry information about the direction of the influence. It would therefore be interesting to consider other measures that are capable of determining the causality of these associations. While we have used the longest available period in our data set (20 years), further investigations may include longer historical data and probe the effects of financial market age and other features that the financial network carries. Similarly, it would be enlightening to create other financial networks with high-frequency intra-day data, or to use a multiscale approach. It is also worth noticing that stock market indices comprise the aggregated prices of several stocks, and it would be fascinating to investigate a similar financial network composed of a large number of individual stocks. We believe these limitations open several possibilities for future research, and we hope they will inspire other studies with a goal to better understand financial data with physics-inspired approaches.

\section*{Methods}

\subsection*{Data}\label{data}

The data set used in our study was obtained from the \textit{Yahoo!} finance historical data API (via the Python module \textit{yfinance}~\cite{yfinance}), the Wall Street Journal market data~\cite{wsj}, and \textit{investing.com} (via the Python module \textit{investpy}~\cite{investpy}). We have first collected the ticker symbol of all 43 major stock market indices (Table~\ref{stab:1}), and next retrieved the adjusted daily closing prices of each one from \textit{Yahoo!} finance in the period from January 1, 2000 to October 31, 2020 (each time series has 5204 data points). The tickers missing or with incomplete data in the \textit{Yahoo!} finance database were then retrieved from the Wall Street Journal markets. For tickers that were not available or remained incomplete with the previous data source, we retrieved data from \textit{investing.com}. Thus, all data necessary to reproduce our findings are freely available.

\subsection*{Permutation entropy}\label{permutation}
The permutation entropy~\cite{bandt2002permutation} is a complexity measure originally proposed for characterizing time series. This measure is calculated from a probability distribution related to local ordering patterns among consecutive time series elements. To define the permutation entropy, we consider a generic time series $\{x_t\}_{t=1,2,\ldots,n}$ and overlapping partitions of length $d>1$ (the embedding dimension) represented by
\begin{equation}
(\vec s\,)\mapsto (x_{s-(d-1)},x_{s-(d-2)},\ldots,x_{s-1},x_s),
\end{equation}
where $s=d,d+1,\ldots,n$. First, for each one of these $(n-d+1)$ partitions, we investigate the $d!$ permutations $\pi = (r_0,r_1,\ldots,r_{d-1})$ of the symbols $(0,1,\ldots,d-1)$ defined by the ordering $x_{s-r_{d-1}}\leq x_{s-r_{d-2}}\leq \ldots \leq x_{s-r_0}$. These permutations symbols represent all $d!$ possible ordering patterns among the $d$ elements of the $(\vec s\,)$ partitions. Next, we estimate the relative frequency of each one of these $d!$ permutation symbols
\begin{equation}
p(\pi_i) = \frac{\mbox{the number of $s$ that has type } \pi_i}{(n-d+1)},
\end{equation}
to define the probability distribution of the ordinal patterns $P=\{p(\pi_i)\}_{i=1,2,\ldots,d!}$. Finally, the permutation entropy~\cite{bandt2002permutation} is simply the normalized Shannon entropy~\cite{shannon1948mathematical} of $P$, that is,
\begin{equation}
H(P)=-\frac{1}{\ln(d!)}\sum_{i=1}^{d!}p(\pi_i)\ln p(\pi_i),
\end{equation}
where $\ln(d!)$ is a normalization constant representing the maximum value of the Shannon entropy. 

The embedding dimension $d$ defining the length of the overlapping partitions is the only parameter of the method. In practice, the choice of $d$ is limited by the time series length $N$ such that the condition $d!\ll n$ must hold to have a reliable estimate of $P=\{p(\pi_i)\}$. We have used $d=4$ in all our results as the length of the sliding window is $500$ trading days. By definition, the normalized permutation entropy is bounded to the interval $0\leq H \leq1$, where values close to the upper bound ($H\approx1$) occur for random time series and lower values of $H$ indicate that the time series exhibits a more complex ordering dynamics. Mainly because of its simplicity, discrimination capabilities, and fast computational evaluation, the permutation entropy framework has successfully been used in many applications~\cite{ribeiro2012complexity,zunino2012review,li2014permutation,sovanovic2016stochastic,sstosic2016investigating,antonelli2017permutation,antonelli2018mammographic,sigaki2018history,sigaki2019estimating}.

\subsection*{Efficiency degree of stock markets}\label{informational}

To define the informational efficiency $H(t)$ used in our study, we first estimate the logarithmic daily price returns (or log-return) series defined by~\cite{mantegna1999introduction}
\begin{equation}
R(t) = \log P(t)-\log P(t-1),
\end{equation}
where $\log P(t)$ and $\log P(t-1)$ are the natural logarithm of the closing indices at time $t$ and $t-1$. Next, we sample the log-return series with a 500-day sliding window that moves ahead one trading day at a time. For each of these time-windows, we calculate the normalized permutation entropy $H(t)$; here, $t$ stands for the last date in the time-windows. This procedure defines a new time series representing the permutation entropy (our measure of informational efficiency) in each window $H(t)$ (Fig.~\ref{fig:1}B). The permutation entropy estimates the degree of randomness in a time series, therefore, entropy values close to one indicate markets at a high informational efficiency state. Conversely, values smaller than one indicate less efficient states of stock markets.

\subsection*{Hierarchical clustering procedure}\label{cluster}
To compute the similarities in the time evolution of the efficiency degree among stock markets, we use the correlation distance matrix
\begin{equation}
d(H_i,H_j) = \sqrt{2(1 - \rho(H_i,H_j))},
\end{equation}
where $\rho(H_i,H_j)$ is the Pearson correlation coefficient between the entropy time series $H_i$ of $i$-th stock market and $H_j$ the same quantity for $j$-th stock market. For the dynamical clustering analysis, $H_i$ and $H_j$ represent the time series segments of stock markets $i$ and $j$ obtained by sampling the efficiency time series with a one-year sliding window. From this correlation distance matrix (Fig.~\ref{fig:2}A), we use the Ward linkage criteria~\cite{hastie2013elements,ward1963hierarchical} to hierarchically cluster the stock markets (dendrogram in Fig.~\ref{fig:2}A). This clustering procedure recursively merges pair of clusters that minimally increase within-cluster variance. 

The threshold distance used to cut the dendrogram and determine the number of clusters is obtained by maximizing the silhouette score~\cite{rousseeuw1987silhouettes}. This coefficient quantifies the consistency of the clustering procedure and is defined by the average value of
\begin{equation}
s_i = \frac{b_i-a_i}{\max(a_i,b_i)}\,,
\end{equation}
where $a_i$ is the cohesion (the average intra-cluster distance) and $b_i$ is the separation (the average nearest-cluster distance) for the $i$-th index. The higher the average value of the silhouette for all indices, the better the cluster configuration. We use the Python module \textit{scikit-learn}~\cite{pedregosa2011scikit} to compute the silhouette scores and the \textit{SciPy}~\cite{virtanen2020scipy} package to compute the correlation distance matrix.

\subsection*{Stochastic block models}
We estimate the modular structure of Fig.~\ref{fig:4}B by using the stochastic block modeling approach~\cite{holland1983stochastic,funke2019stochastic,peixoto2014hierarchical}. This method has the advantage of directly estimating the marginal probabilities that the network is partitioned by a certain number of groups and the probability that a node belongs to a particular group during the inference process. We have tested different stochastic block models (SBM) to fit our network data: usual SBM, degree-corrected SBM (DCSBM), nested SBM, and nested DCSBM. We have considered the best model as the one with the smallest minimal description length~\cite{peixoto2014hierarchical}, based on the statistical evidence that the model is not mistaking stochastic fluctuations for actual modular structure. This means that groups in the network could not have arisen from stochastic fluctuations, as they do in fully random graphs~\cite{guimera2009missing}. We have found that the best model describing our network is the nested stochastic block model without degree correction (nested SBM, see Table~\ref{stab:2}). We have further collected the partitions for 10,000 sweeps of a Metropolis-Hastings acceptance-rejection Markov Chain Monte Carlo~\cite{peixoto2014efficient} with multiple moves to sample hierarchical network partitions, at intervals of 10 sweeps. Using these partitions, we estimate the marginal probability of the number of network modules (Fig.~\ref{sfig:4}), reinforcing that a modular structure with two modules is the most likely structure of our financial network. We have also estimated the marginal probabilities of node membership in our financial network (Fig.~\ref{sfig:5}). The results indicate that the vast majority of markets are almost always assigned to the same partition. All SBM models were implemented with the Python library \textit{graph-tool}~\cite{peixoto_graph-tool_2014}.

\bibliography{collective_efficiency.bib}

\section*{Acknowledgements}

H.V.R. and H.Y.D.S. acknowledge the support of the Coordena\c{c}\~ao de Aperfei\c{c}oamento de Pessoal de N\'ivel Superior (CAPES) and the Conselho Nacional de Desenvolvimento Cient\'ifico e Tecnol\'ogico (CNPq -- Grants 407690/2018-2 and 303121/2018-1). M.P. acknowledges the support of the Slovenian Research Agency (Grants J1-2457 and P1-0403).

\section*{Author contributions statement}

L.G.A.A., H.Y.D.S., M.P., and H.V.R. designed research, performed research, analyzed data, and wrote the paper.

\clearpage

\setcounter{page}{1}
\setcounter{figure}{0}
\rfoot{\small\sffamily\bfseries\thepage/7}%
\makeatletter 
\renewcommand{\thefigure}{S\@arabic\c@figure}
\renewcommand{\thetable}{S\@arabic\c@table}

\begin{center}
\large{Supplementary Information for}\\
\vskip1pc
\large{\bf Collective dynamics of stock market efficiency}\\
\vskip1pc
\normalsize{Luiz G. A. Alves, Higor Y. D. Sigaki, Matja{\v z} Perc, and Haroldo V. Ribeiro}\\
\vskip1pc
\normalsize{Scientific Reports, 2020}\\
\end{center}
\vskip2pc

\begin{table}[!ht]
\begin{center}
\setlength{\tabcolsep}{10pt}
\renewcommand{\arraystretch}{0.8}
\begin{tabular}{cllll}
\toprule
{} &   {\bf Stock market index} & {\bf Location} &     {\bf Ticker symbol} & {\bf Data source}\\
\toprule
1  &                              MERVAL &     Argentina &       \textasciicircum MERV &        Yahoo finance \\
2  &                      ALL ORDINARIES &     Australia &       \textasciicircum AORD &        Yahoo finance \\
3  &                         S\&P/ASX 200 &     Australia &       \textasciicircum AXJO &        Yahoo finance \\
4  &                           ATX Index &       Austria &        \textasciicircum ATX &        Yahoo finance \\
5  &                              BEL 20 &       Belgium &        \textasciicircum BFX &        Yahoo finance \\
6  &                            IBOVESPA &        Brazil &       \textasciicircum BVSP &        Yahoo finance \\
7  &             S\&P/TSX Composite index &        Canada &     \textasciicircum GSPTSE &        Yahoo finance \\
8  &                 SSE Composite Index &         China &   000001.SS &        Yahoo finance \\
9  &                  Shenzhen Component &         China &   399001.SZ &        Yahoo finance \\
10 &                        EURONEXT 100 &        Europe &       \textasciicircum N100 &        Yahoo finance \\
11 &               STOXX Europe 50 Index &        Europe &   \textasciicircum STOXX50E &  Wall Street Journal \\
12 &                              CAC 40 &        France &       \textasciicircum FCHI &        Yahoo finance \\
13 &               DAX PERFORMANCE-INDEX &       Germany &      \textasciicircum GDAXI &        Yahoo finance \\
14 &                     HANG SENG INDEX &     Hong Kong &        \textasciicircum HSI &        Yahoo finance \\
15 &                      S\&P BSE SENSEX &         India &      \textasciicircum BSESN &        Yahoo finance \\
16 &             Jakarta Composite Index &     Indonesia &       \textasciicircum JKSE &        Yahoo finance \\
17 &                         TA-35 Index &        Israel &     TA35.TA &        Yahoo finance \\
18 &         FTSE Italia All-Share Index &         Italy &  FTSEMIB.MI &        Yahoo finance \\
19 &                          Nikkei 225 &         Japan &       \textasciicircum N225 &        Yahoo finance \\
20 &            FTSE Bursa Malaysia KLCI &      Malaysia &       \textasciicircum KLSE &        Yahoo finance \\
21 &                          IPC MEXICO &        Mexico &        \textasciicircum MXX &        Yahoo finance \\
22 &                 Amsterdam AEX Index &   Netherlands &        \textasciicircum AEX &        Yahoo finance \\
23 &           Oslo Bors All Share Index &        Norway &      \textasciicircum OSEAX &  Wall Street Journal \\
24 &             Pakistan Stock Exchange &      Pakistan &        \textasciicircum KSE &        Yahoo finance \\
25 &                          PSEi Index &   Philippines &     PSEI.PS &        Yahoo finance \\
26 &                               WIG20 &        Poland &       WIG20 &        investing.com \\
27 &                   MOEX Russia Index &        Russia &    IMOEX.ME &        investing.com \\
28 &  Russian Trading System (RTS) Index &        Russia &     RTSI.ME &  Wall Street Journal \\
29 &            Tadawul All Shares Index &  Saudi Arabia &    \textasciicircum TASI.SR &        investing.com \\
30 &                           STI Index &     Singapore &        \textasciicircum STI &        Yahoo finance \\
31 &            Top 40 USD Net TRI Index &  South Africa &    \textasciicircum JN0U.JO &  Wall Street Journal \\
32 &               KOSPI Composite Index &   South Korea &       \textasciicircum KS11 &        Yahoo finance \\
33 &                       IBEX 35 Index &         Spain &       \textasciicircum IBEX &        Yahoo finance \\
34 &                  Swiss Market Index &   Switzerland &       \textasciicircum SSMI &        Yahoo finance \\
35 &                 TSEC weighted index &        Taiwan &       \textasciicircum TWII &        Yahoo finance \\
36 &                           SET Index &      Thailand &     \textasciicircum SET.BK &        investing.com \\
37 &                            FTSE 100 &            UK &       \textasciicircum FTSE &        Yahoo finance \\
38 &                              Dow 30 &           USA &        \textasciicircum DJI &        Yahoo finance \\
39 &           NYSE AMEX COMPOSITE INDEX &           USA &        \textasciicircum XAX &        Yahoo finance \\
40 &                 NYSE COMPOSITE (DJ) &           USA &        \textasciicircum NYA &        Yahoo finance \\
41 &                              Nasdaq &           USA &       \textasciicircum IXIC &        Yahoo finance \\
42 &                        Russell 2000 &           USA &        \textasciicircum RUT &        Yahoo finance \\
43 &                             S\&P 500 &           USA &       \textasciicircum GSPC &        Yahoo finance \\
\bottomrule
\end{tabular}
\end{center}
\caption{The 43 major world stock markets used in our study.}\label{stab:1}
\end{table}

\begin{table}[!t]
\begin{center}
\setlength{\tabcolsep}{10pt}
\renewcommand{\arraystretch}{0.8}
\begin{tabular}{lll}
\toprule
{\bf Model} & {\bf Abbreviation} & {\bf Minimal description length} \\
\toprule
Stochastic block model & SBM & 55 \\
Nested degree corrected stochastic block model & Nested DCSBM & 48 \\
Degree corrected stochastic block model & DCSBM & 41 \\
Nested stochastic block model & Nested SBM & 29 \\
\bottomrule
\end{tabular}
\end{center}
\caption{Model selection based on the minimal description length.}\label{stab:2}
\end{table}

\begin{figure}[ht]
\centering
\includegraphics[width=0.8\linewidth]{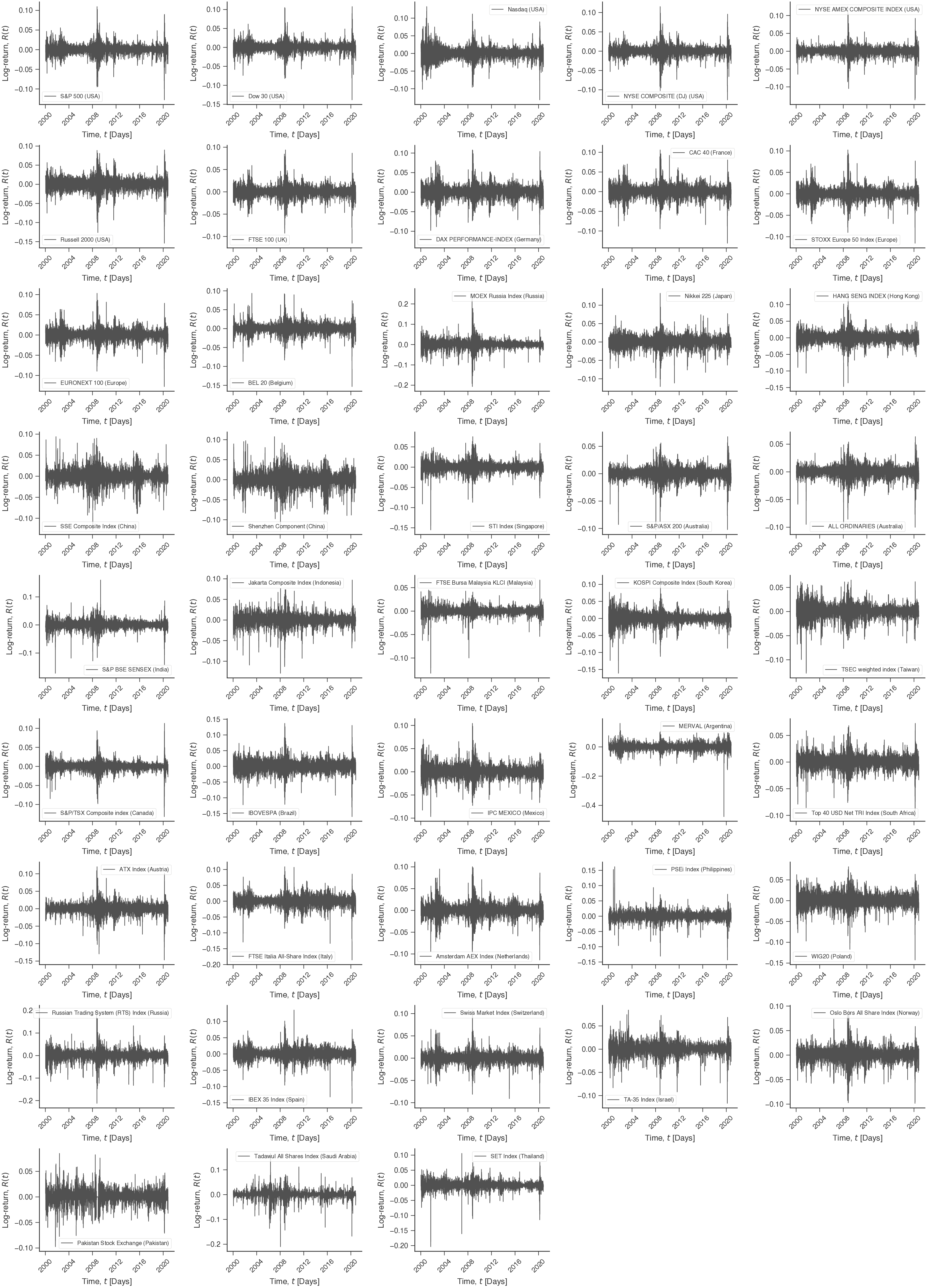}
\caption{Log-returns $R(t)$ time series of the closing prices of the 43 world major stock markets indices from January 1, 2000 to October 31, 2020.}
\label{sfig:1}
\end{figure}

\begin{figure}[ht]
\centering
\includegraphics[width=0.8\linewidth]{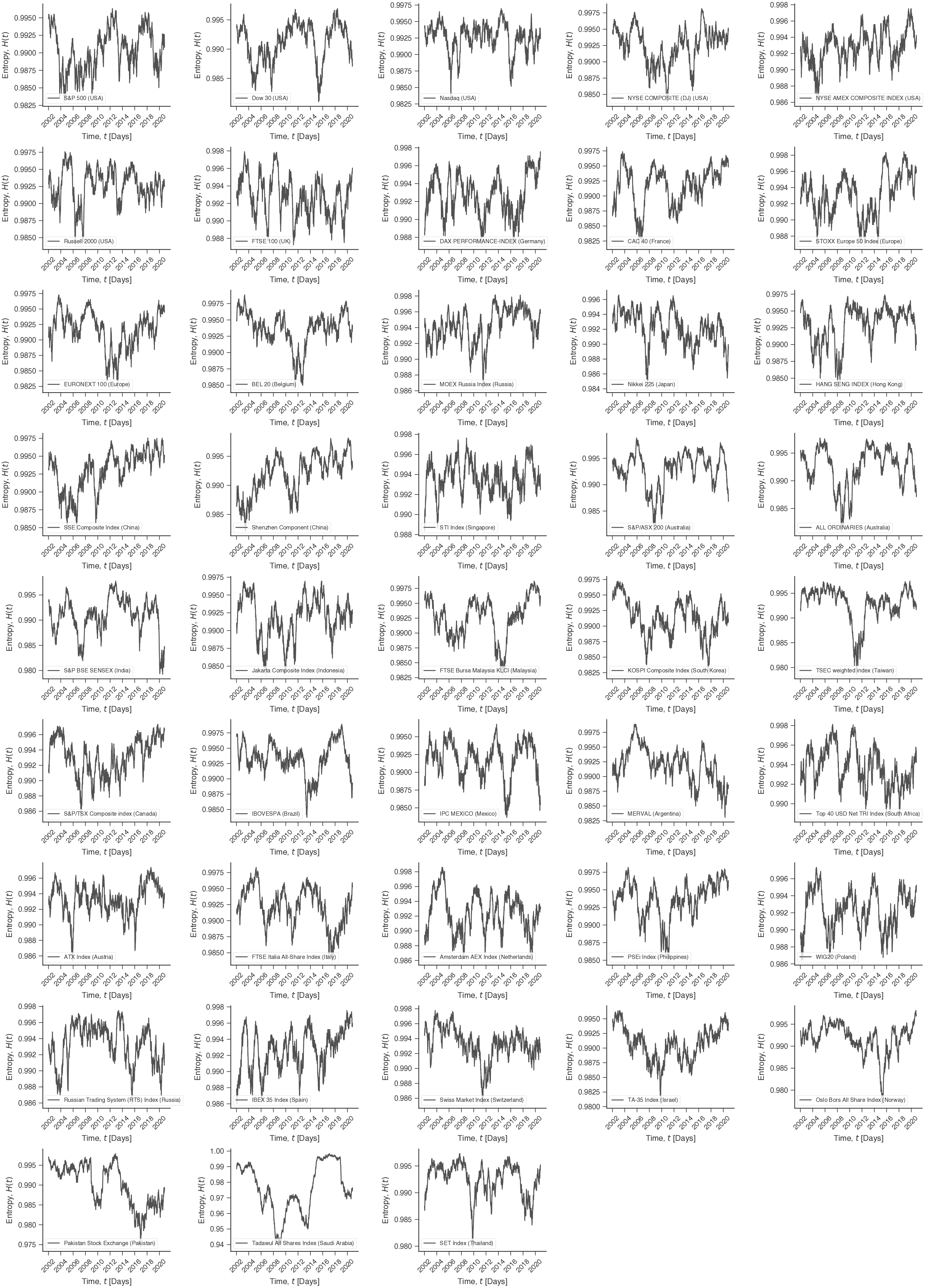}
\caption{Time evolution of the permutation entropy $H(t)$ with embedding dimension $d=4$ (see Methods Section for details) for the 43 major stock markets in our study. }
\label{sfig:2}
\end{figure}

\begin{figure}[ht]
\centering
\includegraphics[width=0.8\linewidth]{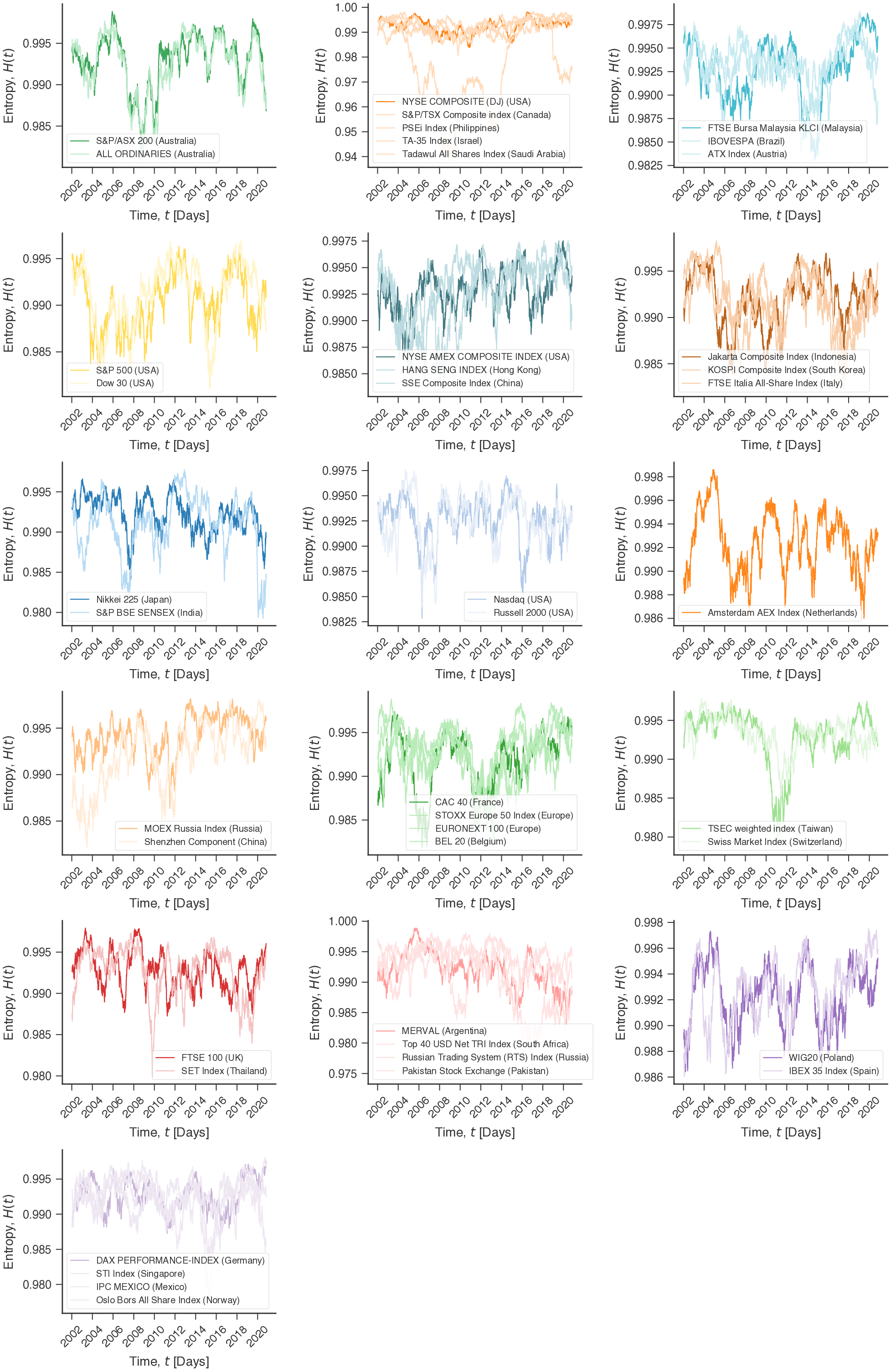}
\caption{Time evolution of the entropy $H(t)$ for the 43 stock markets grouped according to the clusters obtained from the long-term dynamics of $H(t)$.}
\label{sfig:3}
\end{figure}

\begin{figure}[!t]
\centering
\includegraphics[width=0.5\linewidth]{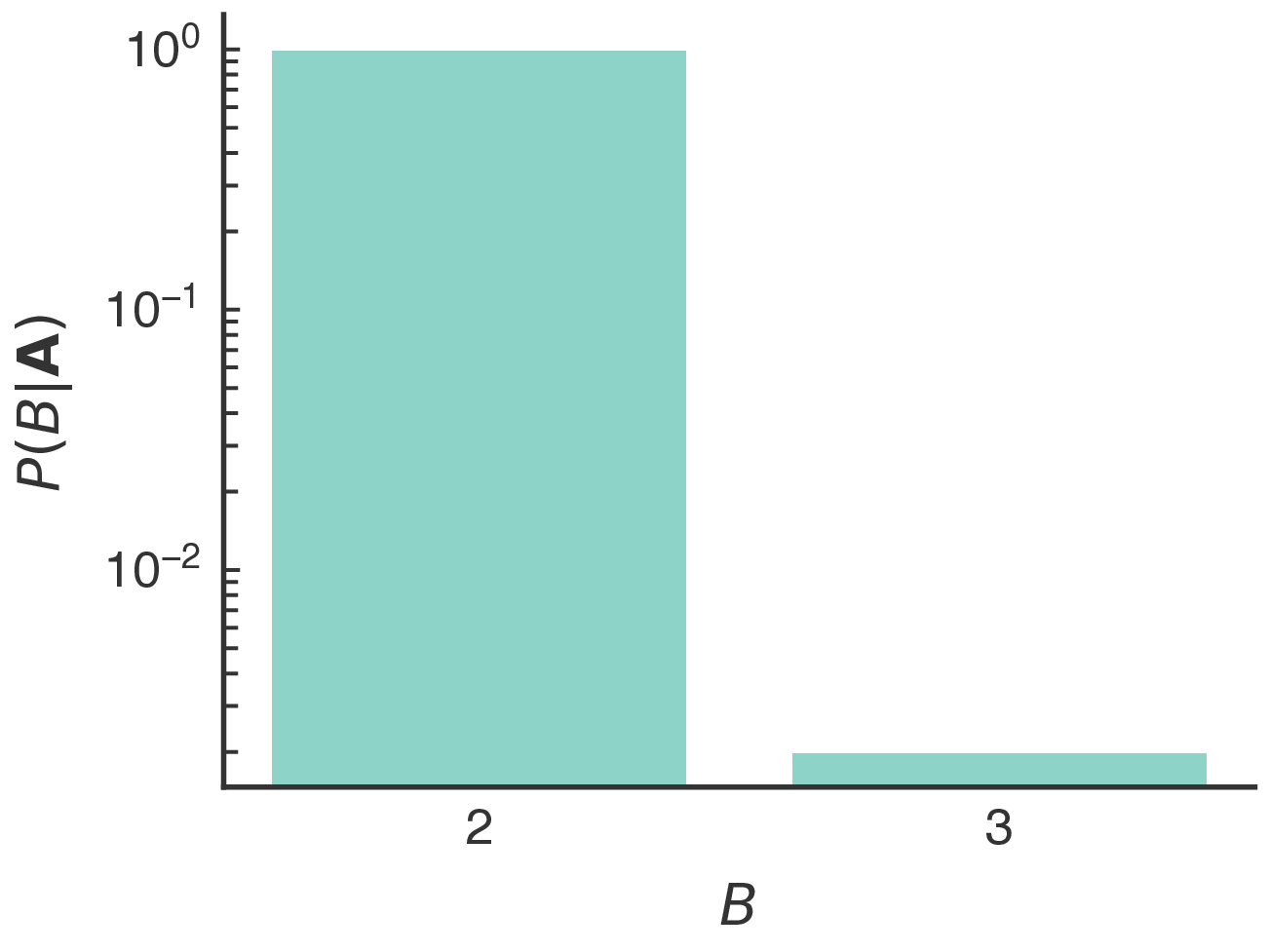}
\caption{Marginal probability of group partition $P(B|{\bf A})$ in our financial network. The bar plot shows the probability density function $P(B|{\bf A})$ that the network ${\bf A}$ is partitioned into $B$ modules. The distribution is concentrated at $B=2$, indicating that two modules is the most likely modular structure for our network. The probabilities were obtained by collecting the number of partitions for 10,000 sweeps of a Metropolis-Hastings acceptance-rejection Markov Chain Monte Carlo with multiple moves to sample hierarchical network partitions, at intervals of 10 sweeps.}
\label{sfig:4}
\end{figure}

\begin{figure}[!ht]
\centering
\includegraphics[width=1\linewidth]{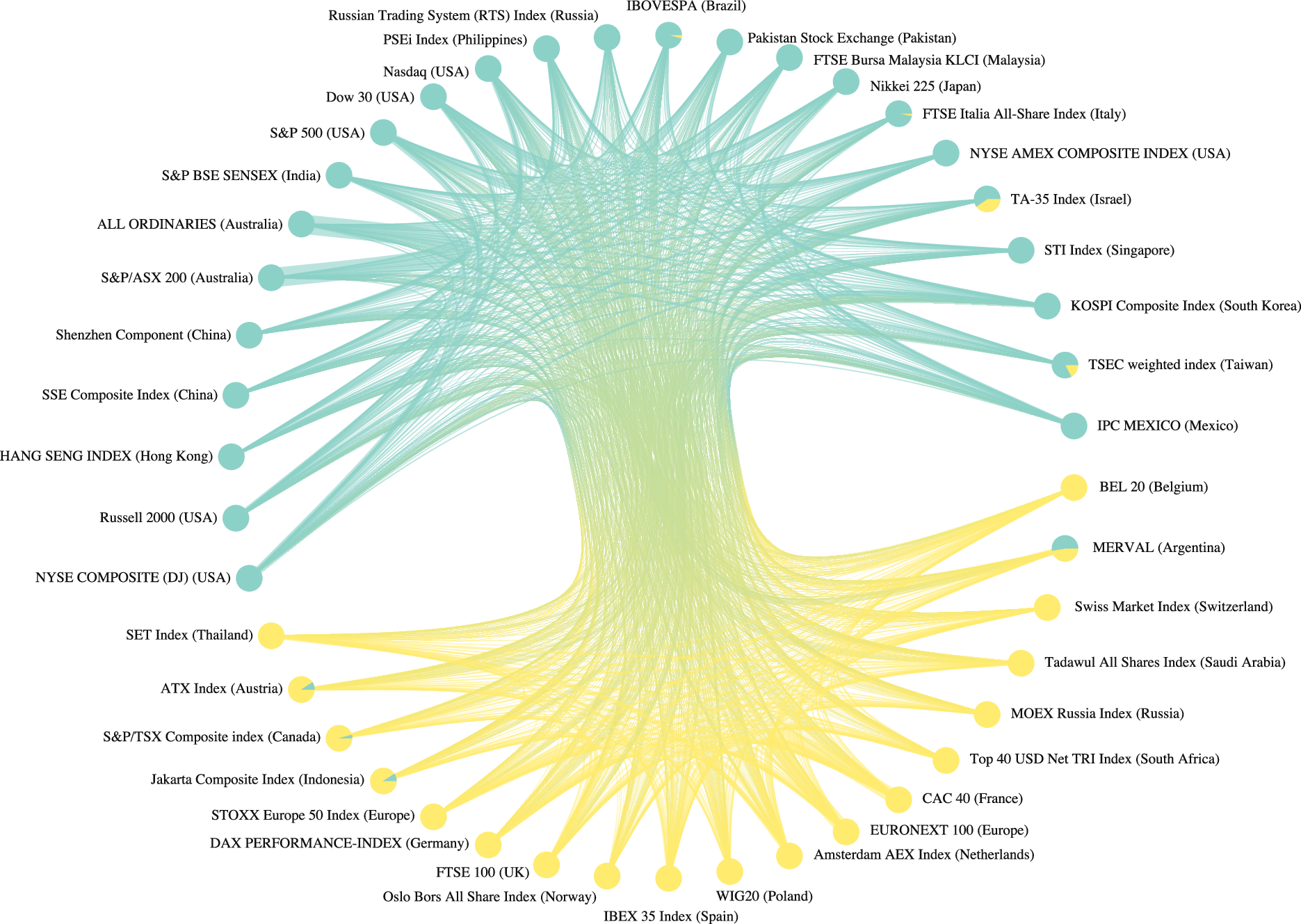}
\caption{Marginal probabilities of node membership in our financial network. In this representation, nodes represent stock markets, and the pie divisions represent the marginal posterior probability that a node belongs to a given group (the two different colors). The probabilities were obtained by collecting the node membership for 10,000 sweeps of a Metropolis-Hastings acceptance-rejection Markov Chain Monte Carlo with multiple moves to sample hierarchical network partitions, at intervals of 10 sweeps. The edges and their weights have the same meaning as those from the network of Fig.~\ref{fig:4} of the main text.}
\label{sfig:5}
\end{figure}

\end{document}